\documentclass[%
 reprint,
 amsmath,amssymb,
 aps,
]{revtex4-2}

\usepackage{graphicx}% Include figure files
\usepackage{dcolumn}% Align table columns on decimal point
\usepackage{bm}% bold math
\usepackage{braket}

\begin{document}

\preprint{APS/123-QED}

\title{Analytical evaluation of the effect of deterministic control error \\  on isolated quantum system}

\author{Kohei Kobayashi}
 \affiliation{Global Research Center for  Quantum Information Science, National Institute of Informatics, 2-1-2 Hitotsubashi,  Chiyoda-ku, Tokyo 101-8340, Japan.}

\date{\today}

\begin{abstract}

We investigate the effect of analog  control errors which deterministically occurrs on isolated quantum dynamics.
Quantum information technologies require careful control for preparing a desired  quantum state used as an information resource.
However, in realistic experiment systems, it is difficult to implement the driving Hamiltonian without analog errors 
and the actual performance of quantum control is far away from the ideal one.
Towards this problem, we derive a lower bound of the overlap between two isolated quantum systems obeying time evolution
in the absence and presence of deterministic control errors.
We demonstrate the effectiveness of the bound through some examples.
Furthermore, by using this bound, we give an analytical estimate on the probability of obtaining the target state under any control errors.

\end{abstract}

\maketitle

%\tableofcontents

\section{Introduction}

Quantum information technologies actively use a time evolution of quantum systems.
To realize an ideal evolution, 
it is needed to implement the Hamiltonian driving the quantum system desirably.
If a quantum system is perfectly isolated from the environments, 
its evolution is simply expressed by the Schr${\rm \ddot{o}}$dinger equation \cite{Nielsen}.
In fact, the scheme of quantum annealing \cite{qa1, qa2, qa3, qa4} or quantum adiabatic computation \cite{qc1, qc2, qc3, qc4} 
are described by it. 

However, physical implementation of quantum control suffer from analog control errors, such as a bias of the magnetic field or detuning caused by the control laser.
In realistic situation, it is difficult to perfectly implement the Hamiltonian without control errors. 
Mainly, analog control errors are divided into deterministic error and stochastic error
(decoherence induced by the interaction with the environment will not be treated in this paper).
In recent years, the influence of stochastic noise on isolated quantum systems has been studied \cite{sn1, sn2}.
The impact of deterministic errors has also been investigated in the literature \cite{dn1, dn2, dn3, dn4}, 
but these are limited to individual issues, and there has been no general consideration so far. 

Inspired by the above facts, we analytically evaluate the deterministic control errors that appear in control Hamiltonians and how this affects quantum state preparation and quantum protocols.
Specifically, we derive the lower bound of the overlap 
between two time-evolved states in isolated quantum system
 with and without control errors  at a given time.
This lower bound is given by the largest eigenvalue of the operator representing the control error.
We demonstrated its effectiveness through simple examples, showing that it gives a better estimation for global control error, rather than collective error.
Furthermore, by using the obtained bound, we provide
an analytical estimation on the probability for getting the target state can be obtained under control errors.

\section{Main result}

We first consider the ideal evolution of the quantum system:
\begin{eqnarray}
\label{id}
 \hbar \frac{d \ket{\psi(t) }  }{dt} 
  = -i\hat{H}(t) \ket{\psi(t)},\ \ 
 \ket{\psi(0)}= \ket{\psi_0}, 
\end{eqnarray}
where $\ket{\psi(t)}$ is the quantum state and $\hat{H}(t)$ is the time-dependent control Hamiltonian.
We assume that $\ket{\psi(t)}$ reaches the target state $\ket{\psi(T)}$ at final time $t=T>0$.
Next, we consider the case where a control error occurs in the Hamiltonian.
In this case, the quantum state $\ket{\phi(t)}$ obeys the following equation:
\begin{eqnarray}
\label{ed}
 \hbar \frac{d\ket{\phi(t)}}{dt} 
 = -i \left( \hat{H}(t)+ \hat{K}(t) \right)\ket{\phi(t) },  
\end{eqnarray}
and $\ket{\phi(0)}= \ket{\psi_0}$.
 $\hat{K}(t)$ is the time-dependent Hermitian operator representing the control error.
What is our interest here is how much $\hat{K}(t)$ affects the state preparation described by Eq. (1).
To analytically evaluate this problem, we define the cost function as follows:
\begin{equation}
\label{fidelity}
P(t)=|\langle \psi(t)|\phi(t)\rangle |,
\end{equation}
which represents the closeness between $\ket{\phi(t)}$ and $\ket{\psi(t)}$.
$P(t)=1$ when $t=0$ and decreases at $t<0$ under $\hat{K}(t)$. 
Under this setting, we present a lower bound of $P(t)$:

{\it Theorem 1.}
The overlap (\ref{fidelity}) has the lower bound at some final time $t=T$ as follows:
\begin{eqnarray}
\label{theorem}
 P(T) \geq 
 P_*= 1- \frac{\alpha^2 T^2}{2 \hbar^2 }, 
\end{eqnarray}
where $\alpha T/\hbar\in[0, \sqrt{2})$.
$\alpha =\overline{ |\lambda_{\rm max}| }$
is the largest eigenvalue $\lambda_{\rm max}$ of $\hat{K}(t)$ 
and time-average $\overline{x}=(1/T)\int^T_0x(t)dt$.
This derivation is based on the method of \cite{Kieu} and given in Appendix A.

The inequality gives a lower bound on the distance between the  states obeying the dynamics in the absence and presence of control error.
Here the important points are listed below.

(i) 
When $T$ is large, it is generally difficult to satisfy the condition $\alpha T/\hbar\in[0, \sqrt{2})$, and this theorem appears to be unusable.
However, it can be possible to reduce the control error by changing the parameters slowly.
In addition, even if the error strength is large,
it may be possible to reduce $T$ by high-speed operations.
Therefore, this condition is not impractical.

(ii)
This result can be extended to the case where the system is subjected to multiple errors $\sum_j\hat{K}_j(t)$.
In this case, $P_*$ is generalized to
\begin{eqnarray}
 P_*= 1- \frac{ (\sum_j \alpha_j T)^2}{2 \hbar^2 }.
\end{eqnarray}
This proof is given in Appendix B.

(iii) In particular, for small $\alpha T<<\hbar$, 
the bound can be approximated as 
\begin{eqnarray}
 P_* \simeq \cos\left( \frac{\alpha T}{\hbar} \right).
\end{eqnarray}
Therefore, the distance between 
 $\ket{\psi(t)}$ and $\ket{\phi(t)}$ diverges more slowly than the cosine function.

\section{Example}

We examine the effectiveness of the bound $P_*$ through some examples.

\subsection{One-qubit}

The first example is the one-qubit system consisting of the exited state $\ket{0}=(1, 0)^\top$
 and the ground state $\ket{1}=(0, 1)^\top$.
We consider the system operators
\begin{equation}
\hat{H} =u \hat{\sigma_y}, \ \ \ \hat{K} =\gamma \hat{\sigma_z},
\end{equation}
where $\hat{\sigma_x}=\ket{0}\bra{1}+\ket{1}\bra{0}$, 
$\hat{\sigma_y}=-i\ket{0}\bra{1}+i\ket{1}\bra{0}$, and  
$\hat{\sigma_z}=\ket{0}\bra{0}-\ket{1}\bra{1}$ are the Pauli matrices.
$\hat{H}$ rotates the state vector along the $y$ axis with frequency $u>0$
and $\hat{K}$ represents the rotation error with strength $\gamma>0$.
In addition, we choose the initial state and the target state as $\ket{\psi_0}=\ket{1}$ and $\ket{\psi(T)}=\ket{0}$, respectively.
In this setting, by solving the equation $\hbar d\ket{\psi(t)}/dt=-iu\hat{\sigma_y}\ket{\psi(t)}$ under $\gamma=0$, 
we can obtain the exact driving time $T=\pi \hbar/(2u)$.
Thus, the lower bound is calculated as follows:

\begin{eqnarray}
\label{boundqubit}
P_* =  1-\frac{ \pi^2 \gamma^2}{ 8u^2},
\end{eqnarray}
where $P_*$ has the meaning if $\gamma/u \in[0, 2\sqrt{2})$.
Figure 1 (a) shows the plots of $P(T)$ and $P_*$ as a function of $\gamma$ when $u=1$ is fixed.
When $\gamma$ is small, $P_*$ works as a tight bound, but as $\gamma$ increases, its tightness becomes weak.
This fact suggests that we should use $P_*$ when $\gamma$ is small compared to $u$. 

For the same setup $(\hat{H}, \ket{\psi_0}, \ket{\psi(T)})$,
we next consider the time-dependent rotation error modeled by 
\begin{equation}
\hat{K}(t) =\gamma\left( \cos(\omega t) \hat{\sigma_x}+ \sin(\omega t)\hat{\sigma_y}\right),
\end{equation}
which is known as a typical control error in quantum computing science. 
$\omega>0$ represents the rotation frequency and  $\alpha=\gamma$.
In this case, the lower bound is given by the same expression as (\ref{boundqubit}) and we find that the difference  between $P(T)$ and $P_*$ is bigger than that of Fig. 1(a) [Fig. 1(b)].
Thereby, we can infer that our bound is more effective for 
errors with simple structures 
than for the one with complicated structures.

\begin{figure}[tb]
\centering
\includegraphics[width=8.5cm]{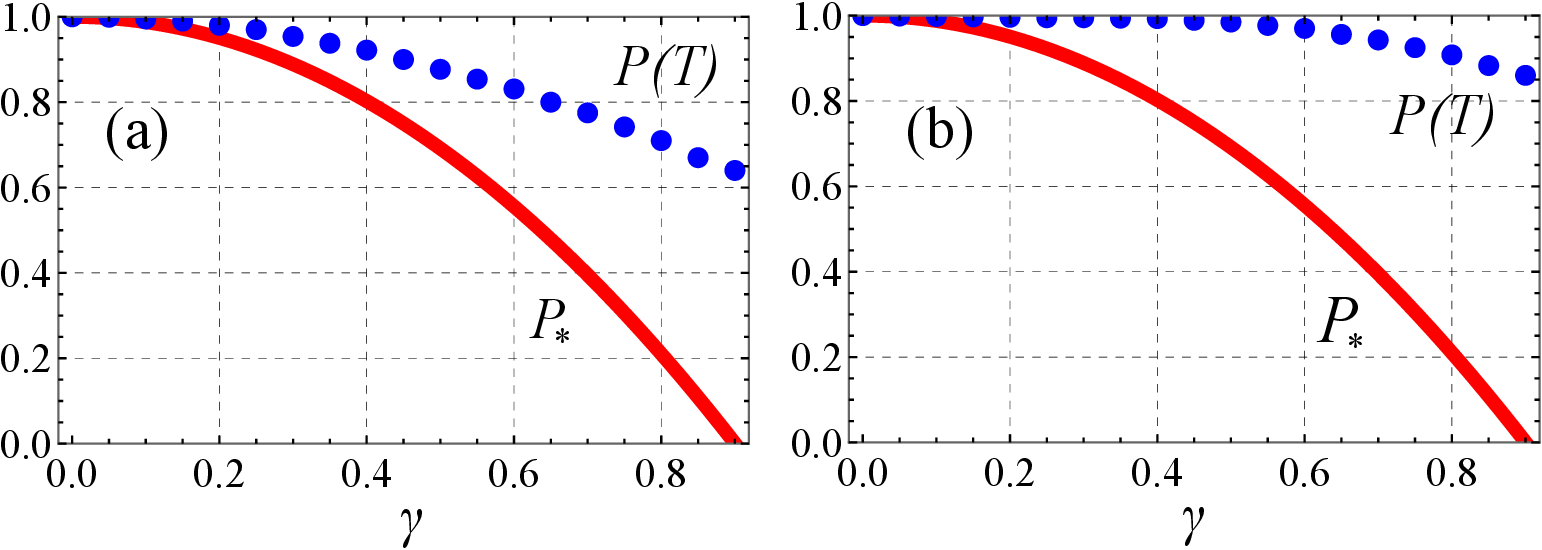}
\caption{  Plots of the simulated values of $P(T)$ and $P_*$ as a function of $\gamma$ for
  (a) time-invariant error and (b) time-dependent error. In both cases, $u=1$ is fixed. }
\end{figure}

\subsection{Two-qubit}

Next, we study the two-qubit system; let us focus on the SWAP operation:
\begin{equation}
\ket{\psi(T)} =\hat{U}_{\rm SWAP} \ket{\psi_0},
\end{equation}
where $\hat{U}_{\rm SWAP}$ is the SWAP gate 

\begin{eqnarray}
 \hat{U}_{\rm SWAP}  =
 \left[  \begin{array}{cccc}
 1 & 0 & 0 & 0   \\ 
0 & 0 & 1 & 0   \\
0& 1 & 0 &0   \\
0 & 0 & 0& 1\end{array}
\right]. 
\end{eqnarray}
It exchanges the states of qubits 1 and 2 and 
plays an important role in quantum
information scenario such as quantum Fourier transform (QFT) \cite{Nielsen}. 
$\hat{U}_{\rm SWAP}$ is generated by the time-independent Hamiltonian
\begin{equation}
\hat{H} =\frac{u}{2}\left( \hat{\sigma}_x\otimes \hat{\sigma}_x  
+ \hat{\sigma}_y\otimes \hat{\sigma}_y +\hat{\sigma}_z\otimes \hat{\sigma}_z \right).
\end{equation}
Assume that the initial state is a product state $\ket{\psi_0}=\ket{0}\otimes\ket{1}$ 
and the target state is 
\begin{equation}
\ket{\psi(T)}= \hat{U}_{\rm SWAP} \ket{\psi_0}=\ket{1}\otimes\ket{0}.
\end{equation}
Then, the control time is calculated as $T=\pi\hbar/(2u)$.
Here we focus on the two types of control errors; 

\begin{eqnarray}
 \hat{K}_G &=& \gamma_G \hat{\sigma}_x\otimes \hat{\sigma}_x,  \\  
\hat{K}^1_C &=&\gamma_{C, 1}\hat{\sigma}_x\otimes \hat{I}, \ \ 
\hat{K}^2_C =\gamma_{C, 2}\hat{I} \otimes \hat{\sigma}_x.   
\end{eqnarray}
$\hat{K}_G$ is a global error acting on the both atoms simultaneously.
On the other hand, $\hat{K}_G$ is a collective error acting on the each atoms, respectively.
Now, to make the expression of $P_*$ same for $\hat{K}_G$ and $\hat{K}_C$, we set $\gamma_G=2\gamma_{C,1}=2\gamma_{C,2}=\gamma$.
Then, we have the following lower  bound:
\begin{equation}
P_*(\hat{K}_G )=P_*(\hat{K}_C )=1-\frac{ \pi^2 \gamma^2}{ 8u^2}.
\end{equation}

Notably, as depicted in Fig. 2(a), $P_*(\hat{K}_G )$ gives a much tighter bound for the simulated values of $P(T)$.
This is thought to be because the error $\hat{\sigma}_x\otimes \hat{\sigma}_x$ is included in $\hat{H}$, resulting in too much control.
Meanwhile, $P_*(\hat{K}_C )$ is clearly weak
compared than $P_*(\hat{K}_G )$.
As shown in this example, there is the case
where $P_*$ functions as a powerful tool for evaluating control performance.

\begin{figure}[tb]
\centering
\includegraphics[width=8.5cm]{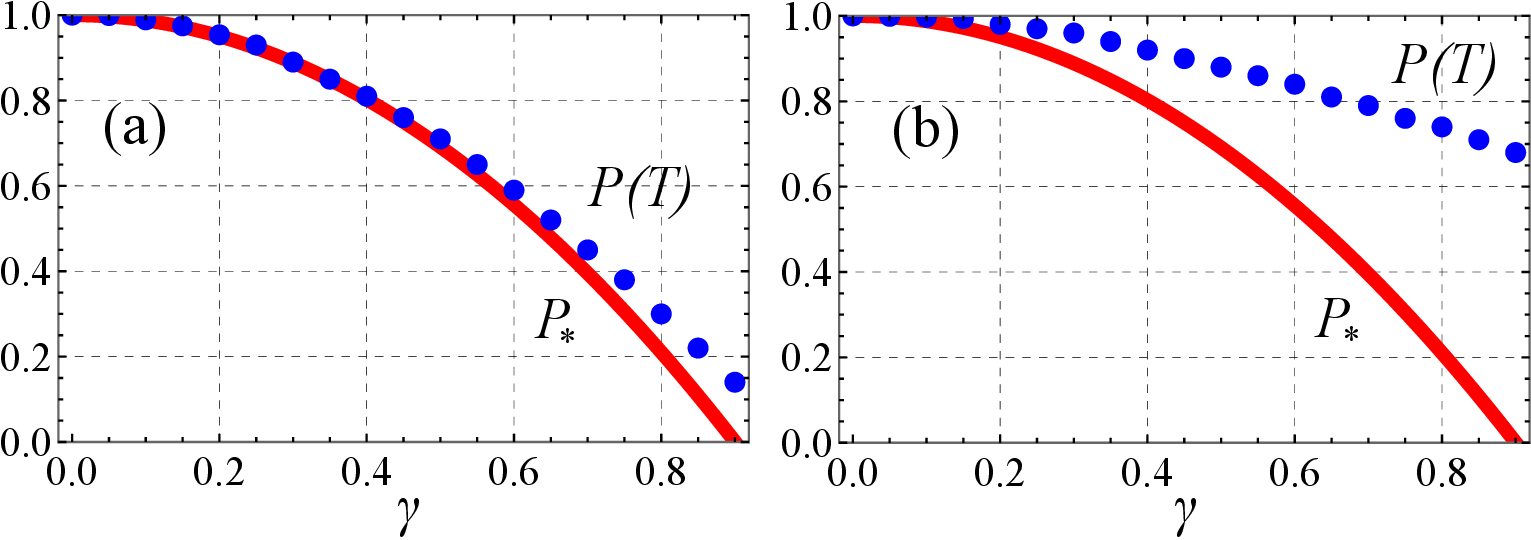}
\caption{  Plots of the simulated values of $P(T)$ and $P_*$ as a function of $\gamma$ for
  (a) global error and (b) collective error. In both cases, $u=1$ is fixed. }
\end{figure}

\section{Application for quantum protocol}

Lastly, we apply our theorem to quantum protocols,
where the computing time is given by a square root of the system size as $T=\mathcal{O}(\sqrt{n})$.
First, as a simple discussion, we set the final overlap to be
\begin{eqnarray}
P(T)=1-\epsilon^2,
\end{eqnarray}
where $0< \epsilon< 1$.
From $P(T)\geq P_*$, $\epsilon$ is upper bounded as
\begin{eqnarray}
\epsilon \lesssim \alpha \times\mathcal{O}(\sqrt{n}).
\end{eqnarray}

Generally, the system size $n$ is sufficiently large, and then, it is necessary to keep the noise strength very small as $\alpha=\mathcal{O}(n^{-1/2})$ in order to achieve high fidelity.
Therefore, when dealing with a large-scale system, powerful techniques for error suppression are required.

As a slightly more concrete example, we consider a quantum search algorithm with control errors.
We aim to find the target state $\ket{m}$ in an unsorted database set of orthogonal states $\{\ket{j},j=1,\cdots, m, \cdots,n\}$. 
Let us approximately write the target state as 
\begin{eqnarray}
\ket{\psi(T)} =\sqrt{1-\epsilon^2}\ket{m} +\sum^n_{j\neq m}p_j\ket{j},
\end{eqnarray}

and further write the another state $\ket{\phi(T)}$ as 
\begin{eqnarray}
\ket{\phi(T)} = \sum^n_{j=1}q_j\ket{j}.
\end{eqnarray}
What we want to know here is how close the actually obtained state  is to the target state.
This problem can be solved by finding a fundamental bound for the probability amplitude $|q_m|$.

For simplicity, we set $q_1=\cdots=q_n=q$;
then, $q=\epsilon/\sqrt{n-1}$.
In this setting, 
we can calculate the upper bound of $P(T)$:

\begin{eqnarray}
P(T) 
&=& \left| \left(\sqrt{1-\epsilon^2}\bra{m} + \sum^n_{j\neq1}p^*_j\bra{j} \right)  \left(\sum^n_{j=1}q_j\ket{j} \right) \right|  \nonumber \\
&=& \left| \sqrt{1-\epsilon^2}q_m +\sum^n_{j\neq m}p^*_jq_j  \right| \nonumber \\
&\leq& |q_m|\sqrt{1-\epsilon^2} +\left|\sum^n_{j\neq m}p^*_jq_j  \right| \nonumber \\
&\leq& |q_m|\sqrt{1-\epsilon^2} + \sum^n_{j\neq m}\left|p^*_jq_j  \right| \nonumber \\
&\leq& |q_m|\sqrt{1-\epsilon^2} 
+ \frac{\epsilon}{\sqrt{n-1}}\sum^n_{j\neq m}\left|q_j  \right|
\nonumber \\
&\leq& |q_m|\sqrt{1-\epsilon^2} 
+ \frac{\epsilon}{\sqrt{n-1}},
\end{eqnarray}
where we used the triangle inequality in the first and second inequality, and the Cauchy-Schwarz inequality in the third inequality.
Combining it and $P_*$, we obtain the lower bound of $|q_m|$:

\begin{equation}
\label{qbound}
|q_m| \geq \frac{1}{\sqrt{1-\epsilon^2}}P_*
- \frac{\epsilon}{\sqrt{1-\epsilon^2}}\frac{1}{\sqrt{n-1}}.
\end{equation}
If $n>>1$ and $\alpha T<<\hbar$ are satisfied, the probability amplitude $|q_m|^2$ is bounded as 
\begin{equation}
|q_m|^2 \gtrsim 
\frac{ 1}{1-\epsilon^2}
{\rm exp}\left\{-\frac{\alpha^2 T^2}{\hbar^2}\right\} .
\end{equation}
Therefore, the probability for getting the target item is  exponentially estimated from below.

\section{Conclusion}
In this paper, we have investigated the effect of deterministic control errors in isolated quantum dynamics, 
by deriving the lower bound for the overlap between two time-evolved states with ideal dynamics and noisy dynamics.
Note that this bound gives a sharper estimation on the achievement of quantum preparation in some cases.
Furthermore, by using the lower bound, we have given a limit on the probability for getting the target state under control errors. 
Finally, we would like to point out that any assumption is not imposed on the time evolution considered in this paper.
Therefore, a future work is to apply our approach to 
practical quantum protocol such as shortcuts to adiabaticity \cite{STA1, STA2}.

\begin{acknowledgments}
This work was supported by MEXT Quantum Leap Flagship Program Grant JPMXS0120351339. 
\end{acknowledgments}

\appendix
\section{Derivation of $P_*$}

From Eqs. (\ref{id}) and (2), we have
\begin{eqnarray}
 \label{ed}
\hbar \frac{d}{dt}\left( \ket{\psi(t) }-\ket{\phi(t)}  \right)
&=&-i \hat{H}(t) \left(\ket{\psi(t) }-\ket{\phi(t)}\right)    \nonumber \\
&+& i \hat{K}(t)\ket{\phi(t) }.   
\end{eqnarray}

Next we consider
\begin{eqnarray}
\label{derive1}
&&  \hbar \frac{d }{dt}\| \ket{\psi(t) }-\ket{\phi(t)} \|^2    \nonumber \\
&=& \hbar \frac{d}{dt} \left\{ \left( \bra{\psi(t) }-\bra{\phi(t)}  \right) \left( \ket{\psi(t) }-\ket{\phi(t)}  \right) \right\} \nonumber \\
 &=&  2 \hbar \Re\left\{ \bra{\psi(t) }-\bra{\phi(t)}  \frac{d}{dt} \left( \ket{\psi(t) }-\ket{\phi(t)}  \right) \right\}.  
\end{eqnarray}

Substituting (\ref{ed}) into  (\ref{derive1}) and using the fact that
$\left( \langle\psi(t)|-\langle \phi(t)|\right)\hat{H}(t)\left( |\psi(t)\rangle-| \phi(t)\rangle \right)$
is a real number because $\hat{H}(t)$ is Hermitian,
\begin{eqnarray}
\label{derive2}
 && 2\hbar \Re\left\{ \bra{\psi(t) }-\bra{\phi(t)}  \frac{d}{dt} \left( \ket{\psi(t) }-\ket{\phi(t)}  \right) \right\}  \nonumber \\
&=& 2 \Im \left\{  \left( \langle\psi(t)|-\langle \phi(t)| \right) \hat{K}(t) |\phi(t) \rangle  \right\}  \nonumber \\
&\leq &  2 \|\ket{\psi(t) }-\ket{\phi(t)} \| \| \hat{K}(t) \ket{\phi(t)}\|  \nonumber \\
&\leq &  2 \|\ket{\psi(t) }-\ket{\phi(t)} \| |\lambda_{\rm max}|,
\end{eqnarray}
where $\lambda$ is the largest eigenvalue of $\hat{K}(t)$.

On the other hand, 
\begin{eqnarray}
\label{derive3}
&&\hbar\frac{d}{dt}\|\ket{\psi(t) }-\ket{\phi(t)} \|^2 \nonumber \\
&=& 2\hbar\|\ket{\psi(t) }-\ket{\phi(t)} \|\frac{d}{dt}\|\ket{\psi(t) }-\ket{\phi(t)} \|.
\end{eqnarray}
Thus, combining  (A3) and (\ref{derive3}),  we have
\begin{eqnarray}
\label{derive4}
\hbar\frac{d}{dt}\|\ket{\psi(t) }-\ket{\phi(t)} \| 
\leq  |\lambda_{\rm max}|,
\end{eqnarray}
 
Then, by integrating this inequality from $t=0$ to $T$, we have
\begin{eqnarray}
\label{Dbound}
 \|\ket{\psi(T) }-\ket{\phi(T)} \| \leq  \ \frac{\alpha T}{\hbar},
\end{eqnarray}
 where we define $\alpha =\overline{ |\lambda_{\rm max}| } $ and time-average $\overline{x}=(1/T)\int^T_0x(t)dt$.
This result gives a meaningful upper bound only 
if $\alpha T/\hbar \in[0, \sqrt{2})$. 
Furthermore, we calculate the lefthand side of (A6) as follows:

\begin{eqnarray}
\|\ket{\psi(t) }-\ket{\phi(t)} \| &=& \sqrt{2-2\Re\{\langle \psi(t)|\phi(t)\rangle \}} \nonumber  \\
&\geq& \sqrt{2(1- P(T))}. 
\end{eqnarray}

Combining this with (\ref{Dbound}), we end up with the following bound
\begin{eqnarray}
 P(T) \geq  
 P_*= 1- \frac{\alpha^2 T^2}{2\hbar^2},
\end{eqnarray}
where $\alpha T/\hbar<[0, \sqrt{2})$.

\section{Generalization of the bound}

We consider a time evolution subjected to the multiple control errors:

\begin{eqnarray}
 \hbar \frac{d\ket{\phi(t)}}{dt} 
 =-i\left( \hat{H}(t) +\sum_j\hat{K}_j(t)  \right)\ket{\phi(t)}.
\end{eqnarray}

Using the result in Appendix A, we can derive the lower bound in the same manner:

\begin{eqnarray}
\hbar \frac{d}{dt}\left( \ket{\psi(t) }-\ket{\phi(t)}  \right)
&=& -i \hat{H}(t) \left(\ket{\psi(t) }-\ket{\phi(t)}\right) \nonumber \\
&\ \ \ +& i\sum_j \hat{K}_j(t)\ket{\phi(t) }.   
\end{eqnarray}

and
\begin{eqnarray}
&& 2\hbar\Re\left\{ \bra{\psi(t) }-\bra{\phi(t)}  \frac{d}{dt} \left( \ket{\psi(t) }-\ket{\phi(t)}  \right) \right\}  \nonumber \\
&=& 2 \Im \left\{  \left( \langle\psi(t)|-\langle \phi(t)| \right) \sum_j\hat{K}_j(t) |\phi(t) \rangle  \right\}  \nonumber \\
&\leq &  2 \|\ket{\psi(t) }-\ket{\phi(t)} \| \times
\| \sum_j \hat{K}_j (t) \ket{\phi(t)}\|  \nonumber \\
&\leq &  2 \|\ket{\psi(t) }-\ket{\phi(t)} \| 
\sum_j  \| \hat{K}_j (t) \ket{\phi(t)}\|  \nonumber \\
&\leq &  2\|\ket{\psi(t) }-\ket{\phi(t)} \|
\sum_j |\lambda_{{\rm max}, j}|.
\end{eqnarray}

Therefore, we have
\begin{eqnarray}
 \|\ket{\psi(T) }-\ket{\phi(T)} \| 
 \leq  \ \frac{ \sum_j \alpha_j T}{\hbar},
\end{eqnarray}

and we obtain the following lower bound:
\begin{eqnarray}
 P(T)\geq  P_*= 1- \frac{( \sum_j \alpha_j T)^2}{2\hbar^2}.
\end{eqnarray}


\begin{thebibliography}{0}%
\makeatletter
\providecommand \@ifxundefined [1]{%
 \@ifx{#1\undefined}
}%
\providecommand \@ifnum [1]{%
 \ifnum #1\expandafter \@firstoftwo
 \else \expandafter \@secondoftwo
 \fi
}%
\providecommand \@ifx [1]{%
 \ifx #1\expandafter \@firstoftwo
 \else \expandafter \@secondoftwo
 \fi
}%
\providecommand \natexlab [1]{#1}%
\providecommand \enquote  [1]{``#1''}%
\providecommand \bibnamefont  [1]{#1}%
\providecommand \bibfnamefont [1]{#1}%
\providecommand \citenamefont [1]{#1}%
\providecommand \href@noop [0]{\@secondoftwo}%
\providecommand \href [0]{\begingroup \@sanitize@url \@href}%
\providecommand \@href[1]{\@@startlink{#1}\@@href}%
\providecommand \@@href[1]{\endgroup#1\@@endlink}%
\providecommand \@sanitize@url [0]{\catcode `\\12\catcode `\$12\catcode `\&12\catcode `\#12\catcode `\^12\catcode `\_12\catcode `\%12\relax}%
\providecommand \@@startlink[1]{}%
\providecommand \@@endlink[0]{}%
\providecommand \url  [0]{\begingroup\@sanitize@url \@url }%
\providecommand \@url [1]{\endgroup\@href {#1}{\urlprefix }}%
\providecommand \urlprefix  [0]{URL }%
\providecommand \Eprint [0]{\href }%
\providecommand \doibase [0]{https://doi.org/}%
\providecommand \selectlanguage [0]{\@gobble}%
\providecommand \bibinfo  [0]{\@secondoftwo}%
\providecommand \bibfield  [0]{\@secondoftwo}%
\providecommand \translation [1]{[#1]}%
\providecommand \BibitemOpen [0]{}%
\providecommand \bibitemStop [0]{}%
\providecommand \bibitemNoStop [0]{.\EOS\space}%
\providecommand \EOS [0]{\spacefactor3000\relax}%
\providecommand \BibitemShut  [1]{\csname bibitem#1\endcsname}%
\let\auto@bib@innerbib\@empty
%</preamble>
\end{thebibliography}%


\begin{thebibliography}{99}

 \bibitem{Nielsen}
M. A. Nielsen and I. L.  Chuang, 
{\it Quantum Computation and Quantum Information} 
(Cambridge University Press, Cambridge, 2010).


\bibitem{qa1}
T. Kadowaki and H. Nishimori, 
Quantum annealing in the transverse Ising model,
Phys. Rev. E {\bf 58}, 5355 (1998).

 
\bibitem{qa2}
J. Brooke, D. Bitko, T. F. Rosenbaum, and G. Aeppli, 
Quantum Annealing of a Disordered Magnet,
 Science {\bf 284}, 779 (1999).

\bibitem{qa3}
J. Brooke, T. F. Rosenbaum, and G. Aeppli, 
Tunable Quantum Tunneling of Magnetic Domain Walls,
 Nature {\bf 413}, 610 (2001). 


\bibitem{qa4}
G. E.  Santro, R. Martonak,   E. Tossati, and R. Car, 
Theory of Quantum Annealing of an Ising Spin Glass,
Science {\bf 295}, 2427 (2002).


\bibitem{qc1}
L. K. Grover, 
Quantum Mechanics Helps in Searching for a Needle in a Haystack,
Phys. Rev. Lett. {\bf 79}, 325 (1997).
	

\bibitem{qc2}
J. Roland and N. J. Cerf, 
Quantum search by local adiabatic evolution,
Phys. Rev. A {\bf 65}, 042308 (2002).


\bibitem{qc3}
D. Aharonov, W. van Dam, J. Kempe, Z. Landau, S. Lloyd, O. Regev,  
Adiabatic Quantum Computation is Equivalent to Standard Quantum Computation,
SIAM J. Comput. {\bf 37}, 166 (2007).


\bibitem{qc4}
A. Mizel, D. A. Lidar, and M. Mitchell, 
Simple Proof of Equivalence between Adiabatic Quantum
Computation and the Circuit Model,
Phys. Rev. Lett. {\bf 99}, 070502 (2007).

\bibitem{sn1}
M. Okuyama, K. Ohki, and  M. Ohzeki,
Threshold theorem in isolated quantum dynamics with stochastic control errors,
Phil. Trans. R. Soc. A. 381, 20210412 (2022).

\bibitem{sn2}
K. Kobayashi,
Control limit for the quantum state preparation under stochastic control errors,
Proc. Roy. Soc. A {\bf 479}, 2277 (2023).

\bibitem{dn1}
J. Roland and  N. J. Cerf,
Noise resistance of adiabatic quantum computation using random matrix theory,
 Phys, Rev. A {\bf 71}, 032330 (2005).

\bibitem{dn2}
S. Mandra, G. Guerreschi, A. A.-Guzik,
Adiabatic quantum optimization in the presence of discrete noise: Reducing the problem dimensionality,
Phys. Rev. A {\bf 92}, 062320 (2015).

\bibitem{dn3}
S. Muthukrishnan, T. Albash, and D. A. Lidar,
Sensitivity of quantum speedup by quantum annealing to a noisy oracle,
Phys. Rev. A {\bf 99}, 032324 (2019).


%\bibitem{Roland}J. Roland and N. J.  Cerf, 
%Quantum circuit implementation of the Hamiltonian versions of Grover's algorithm,
%Phys. Rev. A {\bf 68}, 06231 (2003).

%\bibitem{Hatomura}T. Hatomura,
%Performance evaluation of invariant-based inverse engineering by %quantum speed limits,
%Phys. Rev. A {\bf 106}, L040401 (2021).

\bibitem{dn4} 
T. Albash, V. M-Mayor, T. Hen, 
Analog errors in ising machines. 
 Quant. Sci. Technol. {\bf 4}, 02LT03 (2019).

\bibitem{Kieu}
T. D. Kieu, 
Proc. Roy. Soc. A, {\bf 475}, 20190148 (2019).

\bibitem{STA1}
M. Demirplak and S. A. Rice,
Adiabatic Population Transfer with Control Fields,
J. Phys. Chem. A {\bf 107}, 9937–9945 (2003).

\bibitem{STA2}
X. Chen, A. Ruschhaupt, S. Schmidt, A. del Campo, D. G.-Odelin, and J. G. Muga,
Fast Optimal Frictionless Atom Cooling in Harmonic Traps: Shortcut to Adiabaticity,
Phys. Rev. Lett. {\bf 104}, 063002 (2010).


\end{thebibliography}
\end{document}